\documentclass[rn,11pt]{ucl_document} 
\usepackage{graphicx}
\usepackage{hyperref} 
\usepackage{named}
\usepackage{latexsym}

\begin{document}

\newlength{\halfwidth}
\setlength{\halfwidth}{\textwidth}
\setlength{\halfwidth}{0.5\halfwidth}
\addtolength{\halfwidth}{-2\tabcolsep}

\newlength{\temp}
\newlength{\tempa}
\newlength{\tempb}



\newcommand{\mybowtie}{Bowtie2\ensuremath{^{GP}}}
\newcommand{\ok}{\ensuremath{\surd}}

\title{Which is faster:\ \
  \mybowtie{} $>$ Bowtie $>$ Bowtie2 $>$ BWA}

\date{21 January 2013} 
\documentnumber{13/03}

\author{
{W. B. Langdon} 
}

\maketitle



\begin{abstract}
We have recently used genetic programming to automatically generate an
improved version 
of Langmead's DNA read alignment tool Bowtie2
\cite[Sect.\ 5.3]{Langdon:RN1209}.
We find it runs more than four times faster than 
the Bioinformatics sequencing tool (BWA)
currently used with short next generation 
paired end DNA sequences by the
\href{http://www.ucl.ac.uk/cancer/}{Cancer Institute},
takes less memory
and yet finds similar matches in the human genome.
\\[1em]

\end{abstract}


\section{Introduction}
As part of the 
\href{http://www.cs.ucl.ac.uk/staff/W.Langdon/gismo/}
{Gismo} project
we have used search based software engineering to automatically tailor
a version of the DNA look up tool Bowtie2 \cite{Langmead:2012:nmeth}
which runs considerably faster than the original released code
on 
``single ended'' short (36~bp) DNA sequences produced by the Broad Institute's
Illumina Genome Analyzer~II Solexa scanner.
The multi-objective goals of \mybowtie{}
were to find matches in the human genome faster
without unduly sacrificing the quality of the matches%
\footnote{The $^{GP}$ suffix denotes Bowtie2 was optimised by
genetic programming \cite{poli08:fieldguide}.}.
On out-of-sample Solexa sequences on average it runs more than 70 times
faster than the original release of Bowtie2
and finds very slightly better matches
\cite{Langdon:RN1209}.

While we would normally advocate re-optimising 
the Bowtie2 C++ code for new circumstances,
in order to ease the wide spread up take of \mybowtie{},
we show the original optimised version
can also process DNA sequences from other sources
by applying it to ``double ended'' short DNA sequence
used by the Cancer Institute for human blood studies.

Although the program is identical,
``double ended'' sequences require \mybowtie{}
to combine the results of looking up two DNA sequences
(one from each end of the sequence).
Naturally this combination code 
was not optimised when using the Broad Institute's ``single ended'' data.
Nevertheless \mybowtie{} is able to find high quality matches and
retains some speed advantage over the original released version of
Bowtie2.
Indeed \mybowtie{} on an ACER aspire~5742 
laptop is able to beat 
BWA~\cite{Bioinformatics-2010-Li-589-95}
on our 3~GHz 32~GB server.

There are many Bioinformatics computer based sequencing tools.
In January 2013,
\href{http://en.wikipedia.org/wiki/List_of_sequence_alignment_software}
{Wikipedia}
alone listed more than 140.
\cite{Bioinformatics-2012-Fonseca-3169-77} considered 60 of them.
Bowtie is one of the most widely used and cited
(on average 485 citations per annum%
\footnote{Citation counts from Google Scholar 14 January 2013}).
Langmead rewrote it in C++ to give Bowtie2
(first released 16$^{th}$~October 2011). 
However BWA is also well respected
(108 cites pa)
and is used  by the Cancer Institute.
We compare these three human written DNA sequence tools with
\mybowtie{}
specifically for the Cancer Institute's own data.
For completeness we would have liked to compare against BLAST 
\cite{ncbi_blast}
(44\,454 cites),
which is often taken as the ``gold standard'' 
for Bioinformatics sequence matching,
however it cannot deal with paired end data and,
as we shall see
in the next section,
even treating each end of each DNA sequence pair separately, it
is far too slow for normal use with nextGen sequences.





\section{Method}
\label{sec:method}

We selected uniformly at random one million pairs 
from the 38\,722\,867 produced by the scanner.
(All the pairs have a 36 DNA base sequence at each end.)
We then ran each program 
(with default parameters to generate SAM output) 
on the sample three times on 
our 32~gigabyte Linux server. 
To allow ease of comparison only a single server CPU core was used.
To check for variability this whole procedure was also repeated three
times.

In a similar way we have also tested
BLAST
(version blastn 2.2.25+)
by running it on a random sample of 1000 DNA sequences.
However it was timed out by a 10 minute CPU limit that we imposed.
(The modern alignment tools can process more than 100 times as 
many sequences within ten minutes.
See Table~\ref{tab:summit}.)
Hence Tables~\protect\ref{tab:summit} and~\ref{tab:anal_sam}
refer only to normal paired end runs with
BWA, Bowtie, Bowtie2 and \mybowtie{}.


\section{Results}


BWA finds more matches than the other three tools
(Table~\protect\ref{tab:summit}, column ``\% pairs matched'').
However the difference between BWA and Bowtie2 is only
0.2\% 
and BWA takes more than three times as long.
The fastest program is Bowtie
but it is almost the same speed as \mybowtie{} 
and find 5-6\% fewer matches than the other tools.
\mybowtie{} and Bowtie2 produce very similar matches
but \mybowtie{} is 
26\% 
faster.
%

\begin{table}
\caption{Mean CPU
time taken to process a million paired-end reads
randomly chosen from the
38\,722\,867 
supplied
against the human genome (NCBI release~37 patch~p5).
($\pm$ is shows the observed standard deviation over the $3\times 3$ runs.)
The fourth column is the percentage
of DNA sequences
where the tool reported a suitable match
for both ends.
The next pair of columns were calculated by randomly taking 1000 
of each of the three large samples of
paired end reads
and where the tool reports a match
calculating the Smith-Waterman score for both ends.
This is normalised by summing the scores and dividing through
by the maximum possible score~(72) and expressing this as a percentage.
(With the usual parameters, i.e.~$\mu=0.33$ and $\delta=1.33$,
a single mismatch at one end corresponds to
a normalised score of
98.2). 
\label{tab:summit}
}
\begin{center}
\begin{tabular}{@{}lr@{ $\pm$ }lc@{}r@{ $\pm$ }lc@{}r@{ $\pm$ }lc@{}}
%
%
%
\\
\multicolumn{1}{@{}c}{Tool} & 
\multicolumn{2}{c}{CPU secs} &
{\%}         &\multicolumn{2}{c}{pairs matched} &
{Normalised} &\multicolumn{2}{c}{Smith-Waterman score} &
RAM memory
\\
BWA      &2140 & 55 &&
83.1 & 0.01 && 
98.4 & 3.3   & 
5.3 GBytes
\\
Bowtie   & 490 & 12 &&
77.2 & 0.01 && 
98.7 & 1.9   & 
2.9 GBytes
\\
Bowtie2  & 630 & 17 &&
82.9 & 0.02 && 
98.4 & 2.6   & 
2.2 GBytes
\\
\mybowtie& 500 & 17 &&
82.1 & 0.02 && 
98.5 & 2.5   & 
2.2 GBytes
\\
\end{tabular}
\end{center}
\end{table}

\begin{table}
\caption{Results of statistical comparisons on a random sample of 3000
paired end DNA sequences
(\mbox{$p=0.05$}, sign test,
$\cdot$ indicates difference is not significant).
Left more or better matches.
Right comparison of match quality where both tools report a match.
BWA finds more or better matches.
Whilst Bowtie finds fewer matches
but they are of the same quality
as those also reported by Bowtie2 or \mybowtie{}.
\label{tab:anal_sam}
}
\vspace{2ex}
\begin{tabular}{@{}l|ccc@{}c@{}l|ccc@{}}
%
{\em more matches}
                & Bowtie & Bowtie2 & \mybowtie & 
\mbox{\ \ \ \ \ \ \ \ \ \ }&
{\em better matches}
                & Bowtie & Bowtie2 & \mybowtie\\
\cline{1-4}\cline{6-9}
BWA       &  Yes   & Yes     &  Yes      &&BWA      &  Yes   & Yes     &  Yes      \\
Bowtie    &        & No      &  No       &&Bowtie   &        & $\cdot$ &  $\cdot$  \\
Bowtie2   &        &         &  $\cdot$  &&Bowtie2  &        &         &  $\cdot$  \\

\end{tabular}
\end{table}

\section{Discussion}

Although we do not see the fabulous speed up
we get when  
our own variant of Bowtie2, \mybowtie{},
is used in the way it was optimised for,
it does performs well on 
paired end DNA sequence data.
Although \mybowtie{} found marginally fewer matches but higher quality matches
than Bowtie2,
the differences were not significant in a sample of 3000 paired end
reads
(see Table~\ref{tab:anal_sam}).

\section{Conclusions}
BWA is currently in use by
UCL's Cancer Institute.
However on typical data it is 
{\bf more than four times slower} 
than \mybowtie{}
and yields only 
1\% 
more valid matches,
see Table~\ref{tab:summit}.

\mybowtie{}
is effectively the same speed as Bowtie
and yet finds matches in the human genome in 5\% more cases.
That is, although Bowtie2 was written to give additional functionality
over Bowtie at the expense of run time,
by optimising Bowtie2 to give \mybowtie{},
we have recovered the lost speed and retained the additional functionality.
(Bowtie/\mybowtie{} are the fastest of the five tools tried.
BLAST is by the far slowest, data not shown.)
On the Cancer Institute's paired end DNA sequence data
\mybowtie{} is
26\% 
faster than Bowtie2 from which it was derived.

\subsection*{Acknowledgements}

I would like to thank
\href{http://www.ucl.ac.uk/cancer/medical-genomics/mg_staff#Gareth.Wilson}
{Gareth Wilson}
of the Cancer Institute.


Funded by EPSRC grant
\href{http://gow.epsrc.ac.uk/NGBOViewGrant.aspx?GrantRef=EP/I033688/1}
{EP/I033688/1}.

\newpage

\bibliographystyle{named}
\bibliography{gp-bibliography,references}

\begin{thebibliography}{}

\bibitem[\protect\citeauthoryear{Altschul \bgroup \em et al.\egroup  }{1997}]{ncbi_blast}
\href{http://dx.doi.org/10.1093/nar/25.17.3389}
{Stephen~F. Altschul, Thomas~L. Madden, Alejandro~A. Schaffer, Jinghui Zhang,}
  Zheng Zhang, Webb Miller, and David~J. Lipman.
\newblock Gapped {BLAST} and {PSI-BLAST} a new generation of protein database
  search programs.
\newblock {\em Nucleic Acids Research}, 25(17):3389--3402, 1997.

\bibitem[\protect\citeauthoryear{Fonseca \bgroup \em et al.\egroup  }{2012}]{Bioinformatics-2012-Fonseca-3169-77}
\href{http://dx.doi.org/10.1093/bioinformatics/bts605}
{Nuno~A. Fonseca, Johan Rung, Alvis Brazma, and John~C. Marioni.}
\newblock Tools for mapping high-throughput sequencing data.
\newblock {\em Bioinformatics}, 28(24):3169--3177, 2012.

\bibitem[\protect\citeauthoryear{Langdon and Harman}{2012}]{Langdon:RN1209}
\href{http://www.cs.bham.ac.uk/~wbl/biblio/gp-html/Langdon_RN1209.html}
{William~B. Langdon and Mark Harman.}
\newblock Genetically improving 50000 lines of {C}++.
\newblock Research Note RN/12/09, Department of Computer Science, University
  College London, Gower Street, London WC1E 6BT, UK, 19 September 2012.

\bibitem[\protect\citeauthoryear{Langmead and  Salzberg}{2012}]{Langmead:2012:nmeth}
\href{http://dx.doi.org/10.1038/nmeth.1923}
{Ben Langmead and Steven~L Salzberg.}
\newblock Fast gapped-read alignment with {Bowtie 2}.
\newblock {\em Nature Methods}, 9(4):357--359, 4 March 2012.

\bibitem[\protect\citeauthoryear{Li and  Durbin}{2010}]{Bioinformatics-2010-Li-589-95}
\href{http://dx.doi.org/10.1093/bioinformatics/btp698}
{Heng Li and Richard Durbin.}
\newblock Fast and accurate long-read alignment with {Burrows-Wheeler}
  transform.
\newblock {\em Bioinformatics}, 26(5):589--595, 2010.

\bibitem[\protect\citeauthoryear{Poli \bgroup \em et al.\egroup  }{2008}]{poli08:fieldguide}
\href{http://www.cs.bham.ac.uk/~wbl/biblio/gp-html/poli08_fieldguide.html}
{Riccardo Poli, William~B. Langdon, and Nicholas~Freitag McPhee.}
\newblock {\em A field guide to genetic programming}.
\newblock Published via \texttt{http://lulu.com} and freely available at
  \texttt{http://www.gp-field-guide.org.uk}, 2008.
\newblock (With contributions by J. R. Koza).

\end{thebibliography}

\appendix
\section{Software Versions Used}

\begin{itemize}
\item BWA  \ \  \ \  \ \  \ \  \ \ 0.6.2-r131

\item Bowtie  \ \  \ \  \ \ \ 0.12.7

\item Bowtie~2 \ \ \ \ 2.0.0-beta2

\item \mybowtie{} 2.0.0-beta2 updated by 7~line patch as described 
in technical report \cite{Langdon:RN1209}.
Available via 
\href{http://www.cs.ucl.ac.uk/staff/W.Langdon/ftp/gp-code/bowtie2gp}
{FTP}\@.

\end{itemize}

\section{Seven Line Change to Bowtie2 (2.0.0-beta2)}


\setlength{\tabcolsep}{0.333\tabcolsep}
\begin{tabular}{@{}rlc|c@{}}
\multicolumn{1}{@{}c}{Source file}&
line
     &Original Code&New Code\\ \hline
%
bt2\_io.cpp & 622 
                   & \tt i < offsLenSampled & \tt i < this->\_nPat\\
[2ex]
sa\_rescomb.cpp &  50 
                       & \tt i < satup\_->offs.size() & \tt 0\\
sa\_rescomb.cpp &  69 
                       & \tt j < satup\_->offs.size()\\
[2ex]
aligner\_swsse\_ee\_u8.cpp & 707 & \multicolumn{1}{c|}{\tt vh = \_mm\_max\_epu8(vh, vf);} & \tt vmax = vlo;\\
aligner\_swsse\_ee\_u8.cpp & 766 & \multicolumn{1}{c|}{\tt pvFStore += 4;}\\
aligner\_swsse\_ee\_u8.cpp & 772 & \multicolumn{1}{c|}{\tt \_mm\_store\_si128(pvHStore, vh);} & \tt vh = \_mm\_max\_epu8(vh, vf);\\
aligner\_swsse\_ee\_u8.cpp & 778 & \multicolumn{1}{c|}{\tt ve = \_mm\_max\_epu8(ve, vh);}\\
\\
\end{tabular}
Adapted from \cite[Figure~16]{Langdon:RN1209}.

\end{document}